\documentclass[rnote]{aa} 
\usepackage{graphicx}
\usepackage{txfonts}
\begin{document}
   \title{A compact radio source in the high-redshift soft gamma-ray blazar IGR~J12319--0749}
   \titlerunning{A compact radio source in IGR~J12319--0749}


   \author{S.~Frey\inst{1}
           \and
           Z.~Paragi\inst{2}
	   \and
	   K.\'E.~Gab\'anyi\inst{3}
           \and
	   T.~An\inst{4,5}
	   }

   \institute{F\"OMI Satellite Geodetic Observatory, PO Box 585, H-1592 Budapest, Hungary\\
              \email{frey@sgo.fomi.hu}
         \and
              Joint Institute for VLBI in Europe, Postbus 2, 7990 AA Dwingeloo, The Netherlands\\
              \email{zparagi@jive.nl}
         \and
              Konkoly Observatory, Research Centre for Astronomy and Earth Sciences, Hungarian Academy of Sciences, PO Box 67, H-1525 Budapest, Hungary\\
              \email{gabanyi@konkoly.hu}
         \and
              Shanghai Astronomical Observatory, Chinese Academy of Sciences, 80 Nandan Road, 200030 Shanghai, China\\
              \email{antao@shao.ac.cn}
         \and
              Key Laboratory of Radio Astronomy, Chinese Academy of Sciences, 210008 Nanjing, China}

   \date{Received 22 November 2012; accepted 7 March 2013}

 
  \abstract
   {Blazars are powerful active galactic nuclei (AGNs) radiating prominently in the whole electromagnetic spectrum, from the radio to the X-ray and gamma-ray bands. Their emission is dominated by synchrotron and inverse-Compton radiation from a relativistic jet originating from an accreting central supermassive black hole. The object IGR~J12319$-$0749 has recently been identified as a soft gamma-ray source with the IBIS instrument of the \emph{INTEGRAL} satellite, coincident with a quasar at high redshift ($z$=3.12).}
   {We studied the radio structure of IGR~J12319$-$0749 to strengthen its blazar identification by detecting a compact radio jet on the milli-arcsecond (mas) angular scale, and to measure its astrometric position accurate to mas level.}
   {We used the technique of electronic very long baseline interferometry (e-VLBI) to image IGR~J12319$-$0749 with the European VLBI Network (EVN) at 5~GHz on 2012 June 19.}
   {IGR~J12319$-$0749 (J1231$-$0747) is a compact radio source, practically unresolved on interferometric baselines up to $\sim$136 million wavelengths. The estimated brightness temperature (at least 2$\times$10$^{11}$~K) indicates that the radio emission of its jet is Doppler-boosted. The accurate position of the compact radio source is consistent with the positions measured at higher energies.
}
   {}


   \maketitle
%

\section{Introduction}

\citet{Bass12} identified the soft gamma-ray source \object{IGR~J12319$-$0749} found by the {\em INTEGRAL} satellite as a radio- and X-ray-emitting object. Based on its optical spectrum, the source is a quasar at $z$=3.12 \citep{Mass12}. From the broad emission lines, the mass of the central black hole is estimated to be 2.8$\times$10$^9$~M$_{\odot}$. \citet{Bass12} collected several other pieces of evidence that suggest the source is an extreme blazar, a flat-spectrum radio quasar with powerful jets. If this is the case, the source is the second most distant blazar detected by {\em INTEGRAL} so far. The X-ray source located within the {\em INTEGRAL} error circle has been recently followed up with the {\em Swift} satellite's X-ray telescope (XRT), and UV-optical telescope (UVOT). The X-ray image confirms that the object coincides with a National Radio Astronomy Observatory (NRAO) Very Large Array (VLA) Sky Survey (NVSS) radio source \citep[VLA D configuration, flux density 60.4 mJy at 1.4 GHz]{Cond98}. The radio source is also listed as an unresolved object ($<$5$\arcsec$) in the VLA Faint Images of the Radio Sky at Twenty-cm (FIRST) survey catalogue \citep[][VLA B configuration, 62.9 mJy flux density and 61.0 mJy/beam peak brightness at 1.4 GHz]{Whit97}. No other radio measurements (at frequencies other than 1.4~GHz, and with better resolution) are known. The X-ray flux seems variable over a period of a few months \citep{Bass12}. From the sparsely available, non-simultaneous data and upper limits, \citet{Bass12} tried to reconstruct the spectral energy distribution of IGR~J12319$-$074, and concluded that it is similar to that of another high-redshift blazar, IGR~J22517+2218 \citep[at $z$=3.668;][]{Falc98}.

Here we report on our high-resolution radio interferometric observation of the radio counterpart of IGR~J12319$-$0749, using the technique of very long baseline interferometry (VLBI) with the European VLBI Network (EVN). This is an excellent tool for confirming with imaging that a source is indeed a blazar with compact radio emission on milli-arcsecond (mas) angular scale; VLBI is also capable of determining the accurate astrometric position of a compact radio-emitting object. 

\section{EVN observations and data reduction}

\begin{figure}[!h]
\centering
  \includegraphics[bb=68 169 507 626, height=80mm, angle=270, clip=]{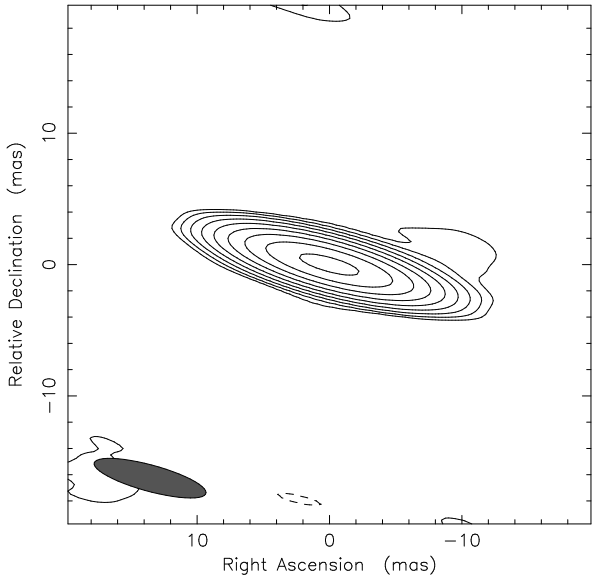}
  \caption{
Naturally-weighted 5-GHz e-EVN image of the quasar J1231$-$0747 (IGR~J12319$-$0749). The lowest contours are drawn at $\pm$0.27~mJy/beam, corresponding to $\sim$3$\sigma$ image noise. The positive contour levels increase by a factor of 2. The peak brightness is 84.1~mJy/beam. The Gaussian restoring beam is 8.8~mas $\times$ 2.0~mas with major axis position angle $75\degr$. The restoring beam (FWHM) is indicated with an ellipse in the lower-left corner. The image is centered on the brightness peak at right ascension $12^{\rm h}31^{\rm m}57\fs68547$ and declination $-7\degr47\arcmin18\farcs0901$.}
  \label{image}
\end{figure}

The EVN observation of IGR~J12319$-$0749 (or J1231$-$0747, the name derived from the more accurate coordinates of the radio counterpart, and used in this paper hereafter) took place on 2012 June 19 at 5~GHz frequency. We utilized the electronic VLBI (e-VLBI) mode \citep{Szom08} where, unlike the conventional VLBI, the signals are not recorded at the telescope sites, but are transmitted to the correlator via wide-band optical fibre networks. The real-time correlation of the data took place at the EVN Data Processor at the Joint Institute for VLBI in Europe (JIVE), Dwingeloo, The Netherlands. At the maximum recording rate of 1024~Mbit~s$^{-1}$, eight antennas of the radio telescope network participated in the experiment: Effelsberg (Germany), Jodrell Bank Mk2 telescope (UK), Medicina, Noto (Italy), Toru\'n (Poland), Onsala (Sweden), Hartebeesthoek (South Africa), and the phased array of the Westerbork Synthesis Radio Telescope (WSRT, The Netherlands). Inter-continental baselines up to the length of $\sim$8100~km were provided by the Hartebeesthoek antenna. Our short exploratory e-EVN experiment (project code RSF06) lasted for 2~h. Eight intermediate frequency channels (IFs) were used in both left and right circular polarisations. The total bandwidth was 128~MHz per polarisation. 

The target source J1231$-$0747 was observed in phase-reference mode \citep[e.g.][]{Beas95}. We did not have prior information about the expected correlated flux density of J1231$-$0747. For a successful detection, we planned sufficiently long coherent integration time to be spent on the source to improve the sensitivity of the observations. Phase-referencing involves regular nodding of the radio telescopes between the target and a nearby bright and compact reference source. We used J1233$-$1025 as the phase-reference calibrator, at $2\fdg65$ separation from our target. The target--reference cycles of $\sim$6~min allowed us to spend $\sim$4.5~min on the target source in each cycle, leading to $\sim$80~min accumulated integration time on J1231$-$0747. The absolute astrometric position of the reference source is listed in the catalogue of the current realisation of the International Celestial Reference Frame \citep[ICRF2,][]{Fey09}. Phase-referencing is suitable for accurately determining the position of the target source with respect to the reference source. 

The US National Radio Astronomy Observatory (NRAO) Astronomical Image Processing System ({\sc AIPS}) was used for the data calibration in a standard way \citep[e.g.][]{Diam95}. The visibility amplitudes were calibrated using the antenna gains, and the system temperatures measured at the antennas during the experiment. Fringe-fitting \citep{Schw83} was performed for the phase-reference calibrator (J1233$-$1025), and the fringe-finder sources (J1125+2610, J1159+2914). The data were then exported to the Caltech {\sc Difmap} package \citep{shep94} for imaging. The conventional mapping procedure involving several iterations of {\sc CLEAN}ing \citep{Hogb74} and phase (then amplitude) self-calibration resulted in the images and brightness distribution models for the calibrators. Overall antenna gain correction factors (typically $\sim$10\% or less) were determined and applied to the visibility amplitudes in {\sc AIPS}. Then fringe-fitting was repeated in {\sc AIPS}, now taking the {\sc CLEAN} component model of the phase-reference calibrator into account. The residual phase corrections resulting from the non-pointlike structure of the phase-reference calibrator were considered this way. The solutions obtained were interpolated and applied to the target source data. Then the visibility data of J1231$-$0747 were also exported to {\sc Difmap} for imaging. The phase-referenced image obtained was used for determining the position of the brightness peak with the {\sc AIPS} task {\sc MAXFIT}. The right ascension $12^{\rm h}31^{\rm m}57\fs68547$ and the declination $-7\degr47\arcmin18\farcs0901$ have accuracies of 0.7~mas and 1~mas, respectively, estimated from the phase-reference calibrator source position accuracy, the target--calibrator separation, the angular resolution of the interferometer array, and the signal-to-noise ratio.

From the phase-referenced data, it turned out that the target source itself is sufficiently bright and compact for fringe-fitting. Therefore, we applied the {\sc AIPS} task {\sc FRING} for J1231$-$0747, as was done earlier for the phase-reference source and the fringe-finders. This way the absolute position information is lost for the target. However, the final naturally-weighted image (Fig.~\ref{image}) obtained in {\sc Difmap} with the standard hybrid-mapping cycles of {\sc CLEAN}ing, and phase and amplitude self-calibration is somewhat more sensitive than the phase-referenced image.

\section{Results and discussion}

Our 5-GHz VLBI image (Fig.~\ref{image}) shows J1231$-$0747 (IGR~J12319$-$0749) as a compact radio source without any extended  feature above the brightness limit of $\sim$0.5~mJy/beam (5$\sigma$) at mas or 10-mas angular scales. The radio source appears practically unresolved with the e-EVN on baselines up to $\sim$136 million wavelengths (M$\lambda$), i.e. from the European antennas to Hartebeesthoek. To physically characterise the source, we used {\sc Difmap} to fit a circular Gaussian brightness distribution model directly to the self-calibrated visibility data. The best-fit model component has 84.6$\pm$4.2~mJy flux density and 0.24$\pm$0.01~mas diameter (full width at half maximum, FWHM). This size can be compared with the minimum resolvable angular size \citep[e.g.][]{Kova05,Loba05} that can be obtained in this VLBI experiment, assuming natural weighting
\begin{equation}
\vartheta_{\rm lim}= b_{\psi} \sqrt{\frac{4 \ln 2}{\pi} \ln{\left(\frac{\rm SNR}{\rm SNR-1}\right)}}.
\end{equation}
Here $b_{\psi}$ is the half-power beam size along a given position angle $\psi$, and SNR the signal-to-noise ratio, i.e. the ratio between the peak brightness and the image noise. 
A source is considered unresolved if its measured visibility amplitude at the longest baseline differs from that of a point source of the same flux density by not more than the 1-$\sigma$ uncertainty.
This method is often applied to determine the minimum resolvable size of compact VLBI-detected components \citep[e.g.][]{Lee08,Savo08,Reyn09,Abdo11,OSul11} 
In our case, SNR=829, and the beam size varies between 2.0~mas and 8.8~mas, corresponding to the FWHM of the elongated elliptical Gaussian restoring beam in the direction of its minor and major axes, respectively (Fig.~\ref{image}). The resulting minimum resolvable angular size is the largest in the direction of the major axis, $\vartheta_{\rm lim, maj}$=0.29~mas. Our fitted model component size (0.24 mas) is similar but somewhat smaller than this value.

With a more conservative approach, we adopt the 3-$\sigma$ image noise level which reduces the signal-to-noise ratio to 276. In this case, the minimum resolvable angular size in the major axis direction becomes $\vartheta_{\rm lim, maj}$=0.50~mas. We can use our modelfit result as the size estimate in the minor axis direction ($\vartheta_{\rm min}$=0.24~mas) since it still exceeds the minimum resolvable size along this position angle.

The measured parameters allow us to estimate the apparent brightness temperature ($T_{\rm b}$) of the compact radio-emitting region \citep{Cond82} in the rest frame of the source
\begin{equation}
T_{\rm b} = 1.22\times 10^{12} (1+z) \frac{S}{\vartheta_{\rm lim, maj} \vartheta_{\rm min} \nu^2},
\end{equation}
where $S$ is the flux density (Jy) and $\nu$ the observing frequency (GHz). For J1231$-$0747, we obtain $T_{\rm b}$=(1.42$\pm$0.09) $\times$ $10^{11}$~K. Because of the unresolved nature of the radio source, this value can be considered a lower limit to the brightness temperature. 

A reasonable assumption about the intrinsic brightness temperature ($T_{\rm b,int}$) of the source would lead to an estimate of the Doppler-boosting factor $\delta$=$T_{\rm b}/T_{\rm b,int}$ \citep{Read94} in the jet. It is customary to assume the equipartition value, $T_{\rm b,int} = T_{\rm eq} \simeq 5\times 10^{10}$~K \citep{Read94,Laht99} as a good approximation of the intrinsic brightness temperature of compact AGN \citep[e.g.][]{Jors06,Hova09,Vere10,Wu12}. This is valid if the energy in the radiating particles is equal to the energy stored in the magnetic field. Departure from equipartition is certainly possible for sources in their maximum brightness state, which could increase $T_{\rm b,int}$ \citep{Homa06}. On the other hand, VLBI studies of large samples found that for the majority of objects the typical $T_{\rm b,int}$ measurements are $\sim$$2-3\times 10^{10}$~K, very close to but somewhat below the canonical equipartition value \citep[e.g.][]{Kell04,Homa06}. 

Bearing in mind that there is always an uncertainty in the determination of Doppler factors from single-epoch brightness temperature measurements, we follow the general practice and adopt $T_{\rm b,int} = 5\times10^{10}$~K for J1231$-$0747. In this case $\delta \ga 2.8$. It can naturally be explained in the framework of the standard orientation-based unified model for radio-loud active galactic nuclei \citep{Urry95}, i.e. with synchrotron emission of the plasma in an approaching relativistic jet pointing close to our line of sight. A rough estimate of the jet inclination angle with respect to the line of sight $\theta$ can be given by assuming a typical bulk Lorentz factor found for quasar jets, $\Gamma$=10 \citep[e.g.][]{Kell04}. Using
\begin{equation}
\cos \theta = \frac{\Gamma - \delta^{-1}}{\sqrt{\Gamma^2 - 1}},
\end{equation}
the jet inclination is $\theta$$\approx$14$\degr$, which becomes smaller if the Doppler factor and/or the Lorentz factor are higher than assumed here. The compact high-resolution radio structure of J1231$-$0747 and its measured brightness temperature are consistent with the VLBI imaging data for most radio-loud quasars at around $z$=3 \citep[e.g.][]{Gurv92,Gurv94,Frey97,Para99,OSul11}. 

Our VLBI experiment provides the first spectral data point available for the source at 5~GHz. The measured $S_{\rm 5}$=84.6$\pm$4.2~mJy is a lower limit to the total flux density of J1231$-$0747 if there is any extended emission around the central compact source that is resolved out by the EVN. This value is higher than the 1.4-GHz flux densities from the NVSS and FIRST surveys (consistently $\sim$60~mJy), which may indicate an inverted power-law spectrum of this quasar in the observed GHz frequency range. Note that the emitted (rest-frame) frequencies are higher by a factor of (1+$z$) because of the expansion of the Universe. In our case, $\nu_{\rm obs}$=5~GHz corresponds to $\nu_{\rm em}$=20.6~GHz. Since the different flux density measurements are made at widely separated epochs, it is also possible that J1231$-$0747 shows significant flux density variations. Both the flat or slightly inverted radio spectrum and the variability are characteristic to compact blazars.

\section{Conclusions}

Our e-EVN observation of the suspected radio counterpart of IGR~J12319$-$0749 (J1231$-$0747), the second-highest redshift soft gamma-ray blazar detected with the {\em INTEGRAL} satellite, provided strong additional support for its blazar nature suggested by \citet{Bass12}. The derived position of the compact radio source is consistent with the less accurate a-priori coordinates taken from the FIRST survey \citep{Whit97} within the errors, thus strengthening the identification of the gamma-ray source with the radio source, the only compact source within the FIRST beam. We found that J1231$-$0747 is a $\sim$85-mJy radio source with sub-mas angular size. The slightly inverted radio spectrum or the flux density variability implied by our measurement at 5~GHz are expected from a blazar. 

With the usual assumption of energy equipartition between the magnetic field and the relativistic particles in the jet \citep{Read94}, we estimated the lower limit to the brightness temperature of J1231$-$0747. It indicates a Doppler-boosted radio jet inclined within $\sim$14$\degr$ to the line of sight. The size of the radio source and the parameters of the jet could be better constrained with future higher-resolution VLBI imaging observations, in particular with longer interferometric baselines in the east--west direction. Simultaneous multi-frequency observations with large radio telescopes or interferometers would reveal the radio part of the SED and determine the turnover frequency in the radio spectrum. Flux density monitoring would shed light on the variability properties of this blazar, possibly providing additional constraints on the physical parameters of its jet. 

From an observational point of view, its sufficiently high brightness and compactness make J1231$-$0747 an ideal new phase-reference calibrator object for any future VLBI experiment studying nearby weak radio sources.

\begin{acknowledgements}
We are grateful to the chair of the EVN Program Committee, Tom Muxlow, for granting us short exploratory e-VLBI observing time. We thank the two anonymous referees for their suggestions. The EVN is a joint facility of European, Chinese, South African, and other radio astronomy institutes funded by their national research councils. e-VLBI research infrastructure in Europe is supported by the European Community's Seventh Framework Programme (FP7/2007-2013) under grant agreement RI-261525 NEXPReS. The research leading to these results has received funding from FP7 under grant agreement No. 283393 (RadioNet3), the Hungarian Scientific Research Fund (OTKA, grant no.\ K104539), and the China--Hungary Collaboration and Exchange Programme by the International Cooperation Bureau of the Chinese Academy of Sciences. T. An is grateful for the financial support from the National Natural Science Foundation of Science and Technology of China (2013CB837901). 
\end{acknowledgements}

\end{document}